\documentclass[conference]{IEEEtran}

\IEEEoverridecommandlockouts

\usepackage{cite}
\usepackage{amsmath,amssymb,amsfonts}
\usepackage{graphicx}
\usepackage{textcomp}
\usepackage{xcolor}
\usepackage{amsmath}
\usepackage{verbatim}
\usepackage{overpic}
\usepackage{algorithm}
\usepackage{algpseudocode}
\usepackage{graphics}
\usepackage{epsfig}
\usepackage{algorithm,algpseudocode}
\def\BibTeX{{\rm B\kern-.05em{\sc i\kern-.025em b}\kern-.08em
    T\kern-.1667em\lower.7ex\hbox{E}\kern-.125emX}}

\begin{document}

\title{Two-Timescale Optimization for Intelligent Reflecting Surface Aided D2D Underlay Communication}

\author{\IEEEauthorblockN{Chang Cai, Huiyuan Yang, Xiaojun Yuan, \IEEEmembership{Senior Member, IEEE}, and Ying-Chang Liang, \IEEEmembership{Fellow, IEEE}}
	\IEEEauthorblockA{\text{Center for Intelligent Networking and Communications, University of Electronic Sci. and Tech. of China, Chengdu, China} \\
		Emails: \{caichang, hyyang\}@std.uestc.edu.cn, xjyuan@uestc.edu.cn, liangyc@ieee.org}
}

\maketitle

\begin{abstract}
	The performance of a device-to-device (D2D) underlay communication system is limited by the co-channel interference between cellular users (CUs) and D2D devices.
	To address this challenge, an intelligent reflecting surface (IRS) aided D2D underlay system is studied in this paper.
	A two-timescale optimization scheme is proposed to reduce the required channel training and feedback overhead, where transmit beamforming at the base station (BS) and power control at the D2D transmitter are adapted to instantaneous effective channel state information (CSI); and the IRS phase shifts are adapted to slow-varying channel mean.
	Based on the two-timescale optimization scheme, we aim to maximize the D2D ergodic rate subject to a given outage probability constrained signal-to-interference-plus-noise ratio (SINR) target for the CU.
	The two-timescale problem is decoupled into two sub-problems, and the two sub-problems are solved iteratively with closed-form expressions.
	Numerical results verify that the two-timescale based optimization performs better than several baselines, and also demonstrate a favorable trade-off between system performance and CSI overhead.
\end{abstract}

%\begin{IEEEkeywords}
%Intelligent reflecting surface (IRS), device-to-device (D2D) communication, two-timescale optimization
%\end{IEEEkeywords}

\section{Introduction}
Device-to-Device (D2D) communication has turned out to be a revolutionary paradigm because of its high spectrum efficiency, low energy consumption and short packet delays \cite{Asadi2014D2DSurvey,Xu2012ResourceAllocation,Wei2012BF_IC}.
The scarcity of spectrum resources forces D2D users to reuse and share underutilized spectrum with cellular systems, inevitably causing Quality-of-Service (QoS) degradation of cellular users (CUs) due to co-channel interference.
Though such degradation can be alleviated to some extent by the use of resource allocation \cite{Xu2012ResourceAllocation}, beamforming and interference cancellation strategies \cite{Wei2012BF_IC}, exploiting new degrees of freedom to further boost the performance of the D2D system remains urgent and promising \cite{Towards2020Rui}.

A recently emerged green and cost-effective technique, named intelligent reflecting surface (IRS), is envisioned to enhance wireless networks by smartly controlling the phase shifts to reconfigure the wireless propagation environment \cite{Towards2020Rui,Rui2019Survey,yuan20206G}.
The signals reflected by the IRS can be constructively superimposed with other links to improve the desired signal power, as well as destructively added to suppress the aforementioned co-channel interference.
In \cite{Towards2020Rui}, the potential energy efficiency improvement was discussed when introducing the IRS to enhance D2D transmission.
An uplink sum rate maximization problem was studied in \cite{Lv2020D2D} for the IRS-assisted D2D communication system in which the D2D devices underlay the CUs. 

The work in \cite{Asadi2014D2DSurvey,Xu2012ResourceAllocation,Wei2012BF_IC,Towards2020Rui,Rui2019Survey,yuan20206G,Lv2020D2D} assumes perfect instantaneous channel state information (CSI) is available at the base station (BS).
In practice, it is generally difficult to support instantaneous CSI acquisition in the D2D underlay system because of the limited coordination between cellular systems and D2D devices \cite{Asadi2014D2DSurvey}.
The deployment of IRS makes the full instantaneous CSI acquisition even impractical, since the channel training and feedback overhead for the IRS-associated links is proportional to the size of the IRS and hence can be prohibitively large.
This motivates us to design an optimization scheme with affordable CSI overhead while still reaping most of the performance gain.

In this paper, we propose a two-timescale optimization scheme for the IRS-assisted D2D underlay transmission.
To be specific, considering that the channel statistical information generally varies slower than the instantaneous CSI \cite{Ma2019ExploitingSCSI,zhao2019TTS}, we design the IRS phase shifts based on the slow-varying channel statistical information with reduced channel training and feedback overhead. %channel training and feedback
Furthermore, for the BS beamforming and the D2D power optimization, only the instantaneous CSI of the compound channel from a transmitter to a receiver, rather than that of the individual transmitter-IRS and IRS-receiver links, is necessary.
Thus, it is reasonable for the BS beamformer and the D2D power to dynamically cater to the instantaneous CSI of the compound channels.
To our best knowledge, this is the first work that exploits two-timescale optimization in the IRS-assisted D2D underlay system.
The main contributions of this paper are summarized as follows.
Firstly, we formulate a novel two-timescale D2D ergodic rate maximization problem with the outage probability constrained cellular transmission QoS requirement.
Secondly, the constrained stochastic successive convex approximation (CSSCA) \cite{Liu2019CSSCA} based large-timescale algorithm and the alternating direction method of multipliers (ADMM) \cite{TaoADMM} based small-timescale algorithm are proposed to iteratively solve the two-timescale stochastic optimization problem with closed-form expressions.
Numerical results verify that the proposed scheme achieves a favorable balance between system performance and CSI overhead.

\begin{figure}
	[!ht]
	\centering
	\includegraphics[width=0.75\columnwidth]{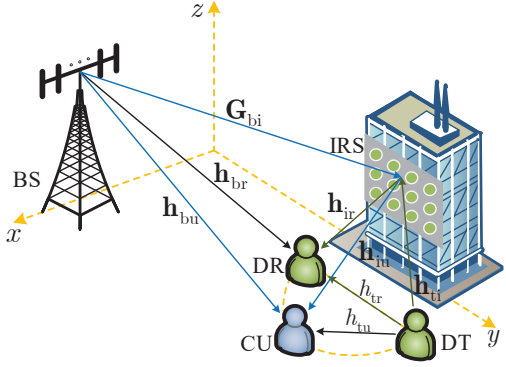}%0.75
	\caption{An IRS-aided D2D underlay communication system.}
	\label{model}
	\vspace{-1.0em}
\end{figure}
\section{System Model}
We consider an IRS-aided D2D communication system which reuses the frequency and time resources assigned to a multiple-input single-output (MISO) downlink cellular transmission, as shown in Fig. \ref{model}.
The system consists of one BS equipped with $M$ antennas, one IRS with $N$ reflecting elements, one single-antenna CU, and a pair of single-antenna D2D users (D2D transmitter and receiver, denoted by DT and DR, respectively).
Let $\mathbf{G}_{\rm bi} \in \mathbb{C}^{N \times M}$, $\mathbf{h}_{\rm iu} \in \mathbb{C}^{1 \times N}$, and $\mathbf{h}_{\rm bu} \in \mathbb{C}^{1 \times M}$ denote respectively the baseband equivalent channels from the BS to the IRS, from the IRS to the CU, and from the BS to the CU.
The counterpart channels from the DT to the IRS, from the IRS to the DR, and from the DT to the DR are represented by $\mathbf{h}_{\rm ti} \in \mathbb{C}^{N \times 1}$, $\mathbf{h}_{\rm ir} \in \mathbb{C}^{1 \times N}$, and ${h}_{\rm tr} \in \mathbb{C}$, respectively.
The baseband equivalent forms of the BS-DR link and the DT-CU link are denoted by $\mathbf{h}_{\rm br} \in \mathbb{C}^{1 \times M}$ and ${h}_{\rm tu} \in \mathbb{C}$, respectively.
We assume a quasi-static flat-fading channel model for all links, and define $\mathbf{H} \overset{\triangle} = \{ \mathbf{G}_{\rm bi}, \mathbf{h}_{\rm iu}, \mathbf{h}_{\rm bu}, \mathbf{h}_{\rm ti}, \mathbf{h}_{\rm ir}, {h}_{\rm tr}, \mathbf{h}_{\rm br}, {h}_{\rm tu} \}$ as the collection of all the above channels.

We assume a Rician fading model for all the channels involved.
For example, the BS-IRS channel $\mathbf{G}_{\rm bi}$ is given by
\begin{equation}
\mathbf{G}_{\rm bi} = \sqrt{\frac{\kappa_{\rm bi}}{\kappa_{\rm bi}+1}} \mathbf{G}_{\rm bi}^{\rm LoS} + \sqrt{\frac{1}{\kappa_{\rm bi}+1}} \mathbf{G}_{\rm bi}^{\rm NLoS},
\end{equation}
where $\kappa_{\rm bi}$ denotes the Rician factor; $\mathbf{G}_{\rm bi}^{\rm LoS}$ is the line-of-sight (LoS) component; and $\mathbf{G}_{\rm bi}^{\rm NLoS}$ is the non-LoS (NLoS) component. 
The BS-IRS channel is dominated by the LoS component because both the BS and the IRS are properly deployed to exploit the LoS path, which remains unchanged for a relatively long time due to their stationary position.
Other channels are similarly modeled with their respective Rician factors.
For low-mobility users, the LoS components of their associated channels may also change slower than the scattering NLoS components\cite{Ma2019ExploitingSCSI,zhao2019TTS}.
The slow-varying property of the LoS components will be exploited in Section \ref{ProFormul} for the IRS phase shifts design.

The IRS reflection amplitudes are restricted to be unit modulus.
Let $\boldsymbol{\Theta} = \mathrm{diag} \left( \theta_1, \dots, \theta_N \right) $ denote the diagonal phase shifts matrix of the IRS elements, where $\theta_n = e^{j\varphi_n}$ with $ \varphi_n \in [ 0,2\pi ) $, $ \forall n = 1, \dots, N$. We also define the vector forms of the phase shifts as $\boldsymbol{\theta} = \left[ \theta_1,\dots, \theta_N \right]^{H} $ and $\boldsymbol{\varphi} = \left[ \varphi_1,\dots, \varphi_N \right]^{T} $.
The signal received at the CU and the DR can be represented respectively as
\begin{equation}
\begin{aligned}
\! y_{\rm u} \! = \! ( \mathbf{h}_{\rm iu} \mathbf{\Theta} \mathbf{G}_{\rm bi} \!  + \! \mathbf{h}_{\rm bu}) \mathbf{w} x_{\rm b} \!  + \!  
(\mathbf{h}_{\rm iu} \mathbf{\Theta} \mathbf{h}_{\rm ti} \!  + \!  {h}_{\rm tu}) \sqrt{p} x_{\rm t}
\!  + \!  n_{\rm u} \label{user receive signal},\!
\end{aligned}
\end{equation}
\begin{equation}
\begin{aligned}
y_{\rm r} \! = \! ( \mathbf{h}_{\rm ir} \mathbf{\Theta} \mathbf{h}_{\rm ti}  \!  + \! {h}_{\rm tr}) \sqrt{p} x_{\rm t} \!  + \! 
( \mathbf{h}_{\rm ir} \mathbf{\Theta} \mathbf{G}_{\rm bi} \!  + \! \mathbf{h}_{\rm br}) \mathbf{w} x_{\rm b}
\!  + \! n_{\rm r} \label{D2 receive signal},
\end{aligned}
\end{equation}
where $\mathbf{w} \in \mathbb{C}^{M \times 1}$ denotes the transmit beamforming vector at the BS; $p$ denotes the transmit power at the DT; $x_{\rm b}$ and $x_{\rm t}$ represent the transmit signal from the BS and the DT, respectively; $n_{\rm u} \sim \mathcal{CN} \left( 0, \sigma_{\rm u}^2 \right)$ and $n_{\rm r} \sim \mathcal{CN} \left( 0, \sigma_{\rm r}^2 \right)$ are the additive white Gaussian noises (AWGNs) at the CU and the DR, respectively. 

Based on \eqref{user receive signal} and \eqref{D2 receive signal}, the received signal-to-interference-plus-noise ratio (SINR) at the CU and the DR can be expressed respectively as
\begin{equation}
\gamma_{\rm u} (\boldsymbol{\varphi},\mathbf{w}, p; \mathbf{H} ) = \frac{\left| ( \mathbf{h}_{\rm iu} \mathbf{\Theta} \mathbf{G}_{\rm bi}  + \mathbf{h}_{\rm bu}) \mathbf{w} \right| ^{2}}
{p \left| \mathbf{h}_{\rm iu} \mathbf{\Theta} \mathbf{h}_{\rm ti} + {h}_{\rm tu} \right|^{2} 
	+ {\sigma_{\rm u}^2} },
\label{user SINR}
\end{equation}
\begin{equation}
\gamma_{\rm r} (\boldsymbol{\varphi},\mathbf{w}, p; \mathbf{H} ) = \frac{p \left| \mathbf{h}_{\rm ir} \mathbf{\Theta} \mathbf{h}_{\rm ti}  + {h}_{\rm tr} \right| ^{2}}
{\left| ( \mathbf{h}_{\rm ir} \mathbf{\Theta} \mathbf{G}_{\rm bi} + \mathbf{h}_{\rm br}) \mathbf{w}\right|^{2}
	+ {\sigma_{\rm r}^2} } 
\label{D2 SINR}.
\end{equation}

Define the effective composite channels from the BS to the CU, from the DT to the CU, from the DT to the DR, and from the BS to the DR respectively as $\mathbf{h}_1 \overset{\triangle} =  \mathbf{h}_{\rm iu} \mathbf{\Theta} \mathbf{G}_{\rm bi}  + \mathbf{h}_{\rm bu}$, ${h}_2 \overset{\triangle} = \mathbf{h}_{\rm iu} \mathbf{\Theta} \mathbf{h}_{\rm ti} + {h}_{\rm tu} $, $ {h}_3 \overset{\triangle} = \mathbf{h}_{\rm ir} \mathbf{\Theta} \mathbf{h}_{\rm ti}  + {h}_{\rm tr} $, and $ \mathbf{h}_4 \overset{\triangle} = \mathbf{h}_{\rm ir} \mathbf{\Theta} \mathbf{G}_{\rm bi} + \mathbf{h}_{\rm br} $.
Denote the collection of the effective channels as $\mathbf{H}_{\rm eff} \overset{\triangle} = \left\{ \mathbf{h}_1, h_2, h_3, \mathbf{h}_4 \right\}$.
From \eqref{user SINR} and \eqref{D2 SINR}, we see that the optimization of $\gamma_{\rm u}$ and $\gamma_{\rm r}$ with respect to (w.r.t.) $\mathbf{w}$ and $p$ only depends on the value of $\mathbf{H}_{\rm eff}$.
This will be exploited in our proposed beamforming and power control design, as detailed in Section \ref{ProFormul}.

\section{Problem Formulation} \label{ProFormul}
Based on the above system model, an immediate design problem is to maximize the achievable rate of the DR over the BS beamformer $\mathbf{w}$, the DT power $p$, and the IRS phase shifts $\boldsymbol{\varphi}$ subject to the QoS constraint of the CU.
Solving this problem, however, requires the full knowledge of the instantaneous CSI $\mathbf{H}$, which may incur unaffordable channel training and feedback overhead.
Furthermore, the limited coordination between the cellular system, the IRS, and the D2D devices makes the approach even impractical.
To avoid this difficulty, we propose a two-timescale optimization scheme to maximize the achievable ergodic rate of the DR instead.
More specifically, we assume that the LoS components of the channels remain unchanged in a relatively long time duration $T_s$ (referred to as the channel statistical coherence time), and that the NLoS components of the channels remain unchanged in a relatively short time duration $T_c$ (referred to as the channel coherence time).
Thus, we use a probability distribution to generate each channel in a time duration of $T_s$, where the LoS component is the mean of the channel (which is a known constant for every time duration of $T_s$) and the elements of the NLoS component are independently and identically drawn from the standard complex Gaussian distribution $\mathcal{CN} \left( 0, 1 \right)$ for every time duration of $T_c$.
Then, we describe the large-timescale optimization and the small-timescale optimization as follows:

\textit{1) Large-timescale optimization:}
To reduce the overhead of instantaneous CSI acquisition and feedback, the IRS phase shifts $\boldsymbol{\varphi}$ are optimized based on the channel statistical information (or more specifically, the channel mean), rather than based on the instantaneous CSI (i.e., the channel realization in each channel coherence time).
Following \cite{zhao2019TTS}, we assume for simplicity that the channel distribution is known at the beginning of each time duration $T_s$.

\textit{2) Small-timescale optimization:}
The BS beamformer $\mathbf{w}$ and the DT power $p$ are optimized based on the instantaneous CSI.
Recall that the optimization of $\mathbf{w}$ and $p$ only needs the instantaneous effective CSI $\mathbf{H}_{\rm eff}$ rather than the full CSI $\mathbf{H}$.
This again reduces the amount of CSI feedback to the BS.

We are now ready to give a rigorous formulation of the two-timescale optimization problem as follows.
For convenience, denote $\Omega \overset{\triangle} = \{\mathbf{w} ~| \left\| \mathbf{w} \right\|^{2} \leq p_0 \}$, $\mathit{\Lambda} \overset{\triangle} = \{p ~| 0 \leq p \leq p_1 \}$, and $\Phi \overset{\triangle} = \{\boldsymbol{\varphi}~| \varphi_n \in [0,2\pi) , \forall n=1,\dots,N \}$ as the feasible sets of the BS beamformer $\mathbf{w}$, the DT power $p$, and the IRS phase shifts $\boldsymbol{\varphi}$, respectively.
Define $\mathbf{w} \left( \cdot \right)$ and $p \left( \cdot \right)$ respectively as the mappings from the sample space of $\mathbf{H}_{\rm eff}$ to $\Omega$ and $\mathit{\Lambda}$.
The DR ergodic rate maximization problem, under the constraint that the SINR outage probability of the CU is below a certain threshold $\epsilon$, can be formulated as
\begin{subequations}\label{OriginalProblem}
	\begin{align}
	 \max_{\boldsymbol{\varphi} \in \Phi, \mathbf{w}\left( \cdot \right), p\left( \cdot \right) } \!\! & \!\! \mathbb{E} \left[  \log_2 \! \Big( 1 \!+ \!\gamma_{\rm r} \big(\boldsymbol{\varphi},\mathbf{w}\left( \mathbf{H}_{\rm eff} \right), p\left( \mathbf{H}_{\rm eff} \right); \mathbf{H} \big)  \Big)  \right] \label{P1_OF} \\
	\operatorname{ s.t. } \quad
	& \!\! \mathbb{P} \big[ \gamma_{\rm u} \big(\boldsymbol{\varphi},\mathbf{w}\left( \mathbf{H}_{\rm eff} \right), p\left( \mathbf{H}_{\rm eff} \right) ; \mathbf{H} \big) \leq \Gamma_{\rm u} \big] \leq \epsilon, \label{P1_C1}
	%&\left| \theta_n \right| =1 , n=1,\dots,N, \label{P1_C4}
	\end{align}
\end{subequations}
where the expectation $\mathbb{E} \left[ \cdot \right]$ and the probability $\mathbb{P} \left[ \cdot \right]$ are taken over the channel distribution.
Problem \eqref{OriginalProblem} is challenging to solve because of the involvement of the mappings $\mathbf{w}\left( \cdot \right)$ and $p\left( \cdot \right)$ in optimization variables, which cannot be expressed in an explicit way.
Moreover, $\boldsymbol{\varphi}, \mathbf{w}\left( \cdot \right)$ and $p\left( \cdot \right)$ are coupled in both the objective function \eqref{P1_OF} and the constraint \eqref{P1_C1}.
Generally, there is no efficient method to solve problem \eqref{OriginalProblem} optimally.
We thus propose an approximate solution to \eqref{OriginalProblem}.
By adopting the primal decomposition technique \cite{boyd2004convex}, problem \eqref{OriginalProblem} can be decomposed into two types of sub-problems, as shown below.

\textit{1) Large-timescale optimization problem} for given the optimal mappings $\mathbf{w}^{\star}\left( \cdot \right)$ and $p^{\star}\left( \cdot \right)$:
\begin{subequations}\label{long-term}
	\begin{align}
	 \max_{\boldsymbol{\varphi} \in \Phi} \quad &  \mathbb{E} \left[ \log_2 \big( 1+\gamma_{\rm r}(\boldsymbol{\varphi}, \mathbf{w}^{\star}, p^{\star};\mathbf{H} ) \big) \right]  \\
	\operatorname{ s.t. } \quad
	& \mathbb{P}\big[ \gamma_{\rm u}(\boldsymbol{\varphi}, \mathbf{w}^{\star}, p^{\star};\mathbf{H} ) \leq \Gamma_{\rm u} \big] \leq \epsilon. \label{P1aC1}
	\end{align}
\end{subequations}

\textit{2) Small-timescale optimization problem} 
%and a series of small-timescale problems with 
for given the optimal $\boldsymbol{\varphi}$ and $\mathbf{H}_{\rm eff} $ (to find $\mathbf{w}^{\star}\left( \cdot \right)$ and $p^{\star}\left( \cdot \right)$):
\begin{subequations}\label{short-term}
	\begin{align}
	\max_{\mathbf{w},p} \quad &  \log_2 \big( 1+\gamma_{\rm r}(\boldsymbol{\varphi},\mathbf{w}, p; \mathbf{H}_{\rm eff}) \big)  \label{small-timescale_obj}\\
	\operatorname{ s.t. } \quad
	& \gamma_{\rm u}(\boldsymbol{\varphi}, \mathbf{w}, p; \mathbf{H}_{\rm eff}) \geq \Gamma_{\rm u} - \delta , \label{short_constraint} \\
	& \left\| \mathbf{w} \right\|^{2} \leq p_0,  \label{w<p0}\\
	& 0 \leq p \leq p_1.
	\end{align}
\end{subequations}
In \eqref{short_constraint}, $\delta$ is appropriately chosen to ensure that \eqref{P1_C1} is met.
It is straightforward to see that when $\delta = 0$, \eqref{short_constraint} is a sufficient condition for \eqref{P1_C1}, but the equality in \eqref{P1_C1} on the right generally does not hold.
As $\delta$ increases, \eqref{P1_C1} becomes tighter.
Thus, we choose $\delta$ such that the right inequality of \eqref{P1_C1} becomes equality.
We emphasize that problem \eqref{OriginalProblem} is not equivalent to its decoupled forms \eqref{long-term} and \eqref{short-term} by noting that \eqref{P1_C1} does not imply \eqref{short_constraint}.
Therefore, solving \eqref{long-term} and \eqref{short-term} generally gives a sub-optimal solution to the original problem \eqref{OriginalProblem}.

\section{Algorithm Design}
In this section, the CSSCA based large-timescale algorithm and the ADMM based small-timescale algorithm are developed to solve the decoupled problems \eqref{long-term} and \eqref{short-term} iteratively.

\subsection{Large-Timescale Optimization} \label{LearningSequence}
The large-timescale optimization is conducted once for every time duration $T_s$ to harvest the performance gain provided by the knowledge of the LoS components.
To tackle the outage probability constraint \eqref{P1aC1}, we transform it into an expectation constrained form as follows.
With the help of the step function $u(x)$, we have $\mathbb{P} \left[ \gamma_{\rm u}  \leq \Gamma_{\rm u} \right] = \mathbb{E}  \left[ u\left( \Gamma_{\rm u} -  \gamma_{\rm u}  \right) \right]  $.
To make the step function differentiable, $u(x)$ is replaced by its smooth approximation $\hat{u}_{\beta}(x) = \frac{1}{1+e^{-\beta x}}$, where $\beta$ is a smooth parameter to control approximation error.
Then, we obtain an approximation of \eqref{P1aC1}:
\begin{align}\label{eq_5}
\mathbb{E} \left[ \hat{u}_{\beta} \big( Q(\boldsymbol{\varphi}, \mathbf{w}^{\star}, p^{\star}; \mathbf{H} ) \big)  \right] - \epsilon \leq 0,
\end{align}
where $Q(\boldsymbol{\varphi}, \mathbf{w}^{\star}, p^{\star}; \mathbf{H} ) = \Gamma_{\rm u} \big(p^{\star} \big| \mathbf{h}_{\rm iu} \boldsymbol{\Theta}   \mathbf{h}_{\rm ti} + {h}_{\rm tu} \big|^{2}+{\sigma_{\rm u}^2}\big) - \big| \big( \mathbf{h}_{\rm iu} \boldsymbol{\Theta} \mathbf{G}_{\rm bi} + \mathbf{h}_{\rm bu} \big) \mathbf{w}^{\star} \big|^2$.

Now, problem \eqref{long-term} can be expressed with expectation forms only:
\begin{subequations} \label{P2}
	\begin{align}
	\min_{\boldsymbol{\varphi} \in \Phi}  \quad & f_0(\boldsymbol{\varphi}) \overset{\triangle} = \mathbb{E} \left[ g_0(\boldsymbol{\varphi}, \mathbf{w}^{\star}, p^{\star}; \mathbf{H} ) \right] \label{std0}\\
	\operatorname{ s.t. } \quad
	&f_1(\boldsymbol{\varphi}) \overset{\triangle} = \mathbb{E} \left[ g_1(\boldsymbol{\varphi}, \mathbf{w}^{\star}, p^{\star}; \mathbf{H} ) \right] \leq 0,\label{std1}
	\end{align}
\end{subequations}
where
\begin{align}
g_0(\boldsymbol{\varphi}, \mathbf{w}^{\star}, p^{\star}; \mathbf{H} ) = -\log_2\big( 1+\gamma_{\rm r}(\boldsymbol{\varphi}, \mathbf{w}^{\star}, p^{\star}; \mathbf{H} ) \big), 
\end{align}
\begin{align}
g_1(\boldsymbol{\varphi}, \mathbf{w}^{\star}, p^{\star}; \mathbf{H} ) =	\hat{u}_{\beta} \big( Q(\boldsymbol{\varphi}, \mathbf{w}^{\star}, p^{\star}; \mathbf{H} ) \big) - \epsilon. 
\end{align}
The CSSCA algorithm \cite{Liu2019CSSCA} is employed to solve the non-convex constrained stochastic optimization problem \eqref{P2} recursively.

Specifically, at each iteration $t$, a new channel sample $\mathbf{H}^t$ is randomly drawn from the known distribution.
For given $\boldsymbol{\varphi}^{t-1}$ (the IRS phase shifts obtained from the $(t-1)$-th iteration) and $\mathbf{H}^t$, the intermediate variables $\mathbf{w}^t$ and $p^t$ are calculated as $\mathbf{w}^t = \mathbf{w}^{\star}(\boldsymbol{\varphi}^{t-1};\mathbf{H}^t)$ and $p^t = p^{\star}(\boldsymbol{\varphi}^{t-1}; \mathbf{H}^t)$ by solving the small-timescale optimization problem \eqref{short-term}.

Then, we approximate the non-convex stochastic functions ${f_0(\boldsymbol{\varphi})}$ and ${f_1(\boldsymbol{\varphi})}$ with the surrogate functions
\begin{equation}
\bar{f}^{t}_m(\boldsymbol{\varphi}) \! = \! c_{m}^{t} + ({\mathbf{f}_m^t})^T(\boldsymbol{\varphi}-\boldsymbol{\varphi}^{t-1}) + \tau_m\| \boldsymbol{\varphi}-\boldsymbol{\varphi}^{t-1} \|^2, m \!= \!0,1,
\label{surrogate_function}
\end{equation}
where $\tau_m >0$ is a constant to ensure convexity; $c_{m}^{t}$ is the sample average approximation of $f_m(\boldsymbol{\varphi})$, i.e.,
\begin{equation}
c_{m}^{t} = \frac{1}{t}\sum_{k=1}^{t}g_m\left(\boldsymbol{\varphi}^{t-1},\mathbf{w}^k, p^k; \mathbf{H}^k \right);
\end{equation}
$\mathbf{f}_m^t = [f_{m,1}^t, \dots, f_{m,N}^t]^T$ is an approximation of the gradient $\nabla {f_m(\boldsymbol{\varphi})}$, which is constructed as a weighted sum of the gradient calculated by the current channel sample and the one from last iteration:
\begin{equation}
\mathbf{f}_m^t = (1-\rho^t)\mathbf{f}_m^{t-1} + \rho^t\nabla_{\boldsymbol{\varphi}}g_m\left(\boldsymbol{\varphi}^{t-1},\mathbf{w}^t,p^t; \mathbf{H}^{t} \right),
\end{equation}
where $\rho^t$ is a decreasing sequence meeting the requirements in \cite{Liu2019CSSCA}, i.e., $\rho^t \rightarrow 0, \sum_{t}\rho^t = \infty, \sum_{t}(\rho^t)^2 < \infty $;
$\nabla_{\boldsymbol{\varphi}}g_m\left(\boldsymbol{\varphi},\mathbf{w},p; \mathbf{H} \right)$ is given in the Appendix for conciseness.

Based on \eqref{surrogate_function}, problem \eqref{P2} is approximated by solving the following convex quadratic optimization problem:
\begin{subequations} \label{P3}
	\begin{align}
	\min_{\boldsymbol{\varphi} \in \Phi} \quad& \bar{f}^{t}_0(\boldsymbol{\varphi}) \\
	\operatorname{ s.t. } \quad
	&\bar{f}^{t}_1(\boldsymbol{\varphi}) \leq 0 \label{feasible_set}.
	\end{align}
\end{subequations}
However, problem \eqref{P3} is infeasible when the channel condition is relatively poor. If \eqref{P3} turns out to be infeasible, the following alternative problem is solved to minimize the constraint \eqref{feasible_set} as the current update: 
\begin{subequations} \label{P3'}
	\begin{align}
	\min_{\boldsymbol{\varphi} \in \Phi} \quad& z \\
	\operatorname{ s.t. } \quad
	&\bar{f}^{t}_1(\boldsymbol{\varphi}) \leq z.
	\end{align}
\end{subequations}
Denote by $\bar{\boldsymbol{\varphi}}^{t}$ the solution of \eqref{P3} or \eqref{P3'} (when \eqref{P3} is infeasible).
\textit{Proposition 1} shows that problem \eqref{P3} can be optimally solved by the Lagrangian dual method. The procedure for solving \eqref{P3'} is similar and hence omitted.

\textit{Proposition 1}\cite[Section \uppercase\expandafter{\romannumeral3}-B]{Liu2018LagrangeProof}:
	problem \eqref{P3} has the following solution given the optimal dual variable $\lambda^{\star}$:
	\begin{align}
	\bar{\varphi}_n^t =  -\frac{b_n(\lambda^{\star})}{2a(\lambda^{\star})} , \forall n = 1, \dots, N, \label{barphi_feasible_update}
	\end{align}
	where $a(\lambda) = \tau_0 + \lambda \tau_1$, $b_n(\lambda) = (f_{0,n}^t-2\tau_0 \varphi_n^{t-1}) + \lambda(f_{1,n}^t-2\tau_1 \varphi_n^{t-1})$.

Finally, $\boldsymbol{\varphi}^{t}$ is updated as
\begin{equation}
\boldsymbol{\varphi}^{t} = (1-\gamma^t)\boldsymbol{\varphi}^{t-1} + \gamma^t \bar{\boldsymbol{\varphi}}^{t}, 
\label{update_phi}
\end{equation}
where $\gamma^t$ is a step size sequence satisfying 1) $\gamma^t \rightarrow 0$, $\sum_{t}\gamma^t = \infty  $, $\sum_{t}(\gamma^t)^2 < \infty$; and 2) $\lim_{t \rightarrow \infty} \frac{\gamma^t}{\rho^t} = 0$ \cite{Liu2019CSSCA}.

The above large-timescale iteration procedure is summarized in Algorithm \ref{Large-timescale}.
\begin{algorithm} [t!]
	\caption{The CSSCA based large-timescale algorithm}\label{Large-timescale}
	\begin{algorithmic}[1]
		\State Initialize $ {\boldsymbol \varphi}^0, t \leftarrow 1  $; and set $\{\rho^{t}\}, \{\gamma^{t}\}$
		\Repeat
		\State Generate a new channel sample $\mathbf{H}^t$, and obtain $\mathbf{w}^t$ and $p^t$ based on the small-timescale algorithm.
		\State Update surrogate functions according to \eqref{surrogate_function}.
		\State Obtain $\bar{\boldsymbol {\varphi}}^t$ according to \eqref{barphi_feasible_update} if problem \eqref{P3} is feasible, otherwise solve problem \eqref{P3'} to obtain $\bar{\boldsymbol \varphi}^t$.
		\State Update ${\boldsymbol {\varphi}}^t$ according to \eqref{update_phi}.
		\State Set $t \leftarrow t+1 $.
		\Until Convergence criteria is met.
	\end{algorithmic}
\end{algorithm}

\subsection{Small-Timescale Optimization}\label{Small-Timescale}
For each effective channel realization, problem \eqref{short-term} is solved based on the alternating optimization (AO) framework.
%instantaneous effective

\textit{1) Optimizing $p$ with fixed $\mathbf{w}$:}
Since \eqref{small-timescale_obj} is monotonically increasing w.r.t. $p$, the optimal $p$ of problem \eqref{short-term} is obtained at the edge of the feasible region, which can be expressed as
\begin{equation}
p = \left\{ \min{ \left[p_1, \frac{1}{\left| {h}_2 \right|^2} \left( \frac{\left| \mathbf{h}_1\mathbf{w} \right|^{2}} {\Gamma_{\rm u} - \delta} - \sigma_{\rm u}^2 \right) \right] }\right\}^{+},   \label{p_update}
\end{equation}
where $\left\{ \cdot \right\}^+$ denotes $\max{ \left\{ \cdot,0 \right\}}$.

\textit{2) Optimizing $\mathbf{w}$ with fixed $p$:}
On the other hand, for any given $p$, problem \eqref{short-term} is a non-convex quadratically constrained quadratic program (QCQP) problem w.r.t. $\mathbf{w}$.
This can be solved by the semi-definite relaxation (SDR) method with complexity of $\mathcal{O}(M^6)$.
To avoid the undesirable computation cost, we adopt the low complexity ADMM algorithm \cite{TaoADMM} to solve problem \eqref{short-term} iteratively with closed-form solutions.

Note that the non-convexity of problem \eqref{short-term} is introduced only by \eqref{short_constraint}, i.e., the QoS constraint of the CU.
This observation motivates us to transform problem \eqref{short-term} (with fixed $p$) into a sequence of convex sub-problems via convex-concave procedure \cite{lanckriet2009convergence}.
Let $\mathbf{w}^{j-1}$ denote the BS beamformer obtained from the $(j-1)$-th iteration.
Then, at the $j$-th iteration, by replacing the concave part of \eqref{short_constraint} with its first-order Taylor expansion at a given point $\mathbf{w}^{j-1}$, we need to solve
\begin{subequations}\label{linear}
	\begin{align}
	 \min_{\mathbf{w}}  \quad & \left| \mathbf{h}_4\mathbf{w} \right|^{2} \\
	\operatorname{ s.t. } \quad
	& 2\Re \{ (\mathbf{w}^{j-1})^H  \mathbf{h}^{H}_1 \mathbf{h}_1 \mathbf{w} \} - \left| \mathbf{h}_1\mathbf{w}^{j-1} \right|^{2} \geq \zeta, \label{Taylor} \\
	& \eqref{w<p0}, \nonumber
	\end{align}
\end{subequations}
where $\zeta = (\Gamma_{\rm u} - \delta)(p\left| {h}_2 \right|^{2}+{\sigma^2_{\rm u}})$.
From \cite{lanckriet2009convergence}, this iterative procedure is guaranteed to converge to a stationary point of the original problem \eqref{short-term}.

For the first step of the ADMM algorithm, we introduce two auxiliary variables $u$ and $\mathbf{v}$ to recast problem \eqref{linear} as
\begin{subequations} \label{P4}
	\begin{align}
	\min_{\mathbf{w},u,\mathbf{v}}  \quad & \left| \mathbf{h}_4\mathbf{w} \right|^{2} \\
	\operatorname{ s.t. } \quad
	&u - \mathbf{h}_1 \mathbf{w} =0, \label{u}\\
	&\mathbf{v} - \mathbf{w} = \mathbf{0},\label{v}\\
	& 2\Re \{ (\mathbf{w}^{j-1})^H \mathbf{h}^{H}_1 u \} - \left| \mathbf{h}_1\mathbf{w}^{j-1} \right|^{2} \geq \zeta, \label{feasible_set1} \\
	&\left\| \mathbf{v} \right\|^{2} \leq p_0. \label{feasible_set2}
	\end{align}
\end{subequations}
Define the feasible region of \eqref{feasible_set1} as $\mathcal{U}$, whose indicator function is given by $\mathbb{I}_{\mathcal{U}}(u)$ (i.e., $\mathbb{I}_{\mathcal{U}}(u) = 0$ if \eqref{feasible_set1} satisfies, otherwise $\mathbb{I}_{\mathcal{U}}(u) = \infty$). 
Similarly, the indicator function of \eqref{feasible_set2} is given by $\mathbb{I}_{\mathcal{V}}(\mathbf{v})$.
Then, we obtain the equivalent ADMM form of problem \eqref{P4} as follows:
%the convex sub-problem at each iteration is reformulated as the ADMM form as
\begin{subequations} \label{P4'}
	\begin{align}
	\min_{\mathbf{w},u,\mathbf{v}}  \quad & \left| \mathbf{h}_4\mathbf{w} \right|^{2} + \mathbb{I}_{\mathcal{U}}(u) + \mathbb{I}_{\mathcal{V}}(\mathbf{v}) \\
	\operatorname{ s.t. } \quad
	&\eqref{u},\eqref{v}. \nonumber
	\end{align}
\end{subequations}
The augmented Lagrangian problem of \eqref{P4'} is
\begin{align}
\mathcal{L}_{\rho} \left( u,\mathbf{v},\mathbf{w},y,\mathbf{z} \right)
= &\left| \mathbf{h}_4\mathbf{w} \right|^{2} + \mathbb{I}_{\mathcal{U}}(u) + \mathbb{I}_{\mathcal{V}}(\mathbf{v})\nonumber + \\
& \frac{\rho}{2} \left| u \! - \! \mathbf{h}_1\mathbf{w} \! + \! y \right|^{2}
+ \frac{\rho}{2} \left\| \mathbf{v}\! - \!\mathbf{w}\! +\! \mathbf{z} \right\|^{2}\!, \label{lagrangian}
\end{align}
where $\rho>0$ is the penalty parameter; $y$ and $\mathbf{z}$ are the dual variables associated with \eqref{u} and \eqref{v}, respectively.

Now, the auxiliary variables $u$, $\mathbf{v}$ and the BS beamformer $\mathbf{w}$ are updated alternatively by minimizing \eqref{lagrangian}.
For given $\mathbf{w}^{j-1}$ and $y^{j-1}$ at the $j$-th iteration, $u^{j}$ is given by
\begin{align}
u^{j} = (\pi^{\star}+1)(\mathbf{h}_1\mathbf{w}^{j-1}) - y^{j-1}, \label{u_update_final}
\end{align}
where $\pi^{\star}$ is the optimal dual variable of the corresponding Lagrangian problem for solving $u$ and the duality gap is zero.
It can also be verified that for given $\mathbf{w}^{j-1}$ and $\mathbf{z}^{j-1}$ at the $j$-th iteration, $\mathbf{v}^{j}$ is updated as
\begin{align}
\mathbf{v}^{j}
= \min \left\{ \frac{\sqrt{p_0}}{\left\|\mathbf{w}^{j-1} - \mathbf{z}^{j-1} \right\| }  ,
1  \right\} \left( \mathbf{w}^{j-1} - \mathbf{z}^{j-1} \right). \label{v_update_final} 
\end{align}
The update of $\mathbf{w}^j$ can be derived as
\begin{align}
\mathbf{w}^{j}
=& \left( 2\mathbf{h}^{H}_4\mathbf{h}_4 + \rho\mathbf{h}^{H}_1\mathbf{h}_1 + \rho \mathbf{I}_M \right)^{-1} \nonumber\\
&\left( \rho \left(u^{j}+y^{j-1}\right)\mathbf{h}^H_1 + \rho \left( \mathbf{v}^{j} + \mathbf{z}^{j-1} \right) \right) , \label{w_update2}
\end{align}
where $\mathbf{I}_M$ denotes the $M \times M$ identity matrix.

\begin{algorithm} [t!]
	\caption{The ADMM based small-timescale algorithm}\label{Algorithm2}
	\begin{algorithmic}[1]
		\State Initialize $\mathbf{w}^0 \leftarrow \sqrt{p_0} \frac{\mathbf{h}^{H}_1}{\left\| \mathbf{h}_1 \right\|} $, $y^0 \leftarrow 0$, $\mathbf{z}^0 \leftarrow \mathbf{0}$; $i \leftarrow 1$, $j \leftarrow 1$; and set the penalty parameter $\rho$.
		\Repeat
		\State Update $p$ according to \eqref{p_update}.
		\Repeat
		\State Update $u,\mathbf{v}, \mathbf{w}$ according to \eqref{u_update_final}, \eqref{v_update_final} and \eqref{w_update2}.
		%\State Update $\mathbf{w}$ according to \eqref{w_update2}.
		\State Update $y, \mathbf{z}$ according to \eqref{y_update} and \eqref{z_update}.
		\State Set $j \leftarrow j+1 $.
		\Until Convergence criteria is met.
		\State Set $i \leftarrow i+1 $.
		\Until Convergence criteria is met.
	\end{algorithmic}
\end{algorithm}

Finally, we update the dual variables $y$ and $\mathbf{z}$ as
\begin{equation}
y^{j} = y^{j-1} + (u^{j} - \mathbf{h}_1 \mathbf{w}^{j}), \label{y_update}
\end{equation}
\begin{equation}
\mathbf{z}^{j} = \mathbf{z}^{j-1} + (\mathbf{v}^{j} - \mathbf{w}^{j}). \label{z_update}
\end{equation}

The small-timescale optimization procedure is shown in Algorithm \ref{Algorithm2}.
We note that the small-timescale algorithm is not only executed for each short time duration $T_c$, but also called by the large-timescale algorithm (in Line 3 of Algorithm \ref{Large-timescale}).

\subsection{Complexity Analysis}
For the large-timescale optimization algorithm, the ellipsoid method \cite{boyd2004convex} is adopted to find the optimal dual variable $\lambda^{\star}$ in \eqref{barphi_feasible_update} with complexity of $\mathcal{O} (N\log(\frac{1}{\epsilon_e}))$, where $\epsilon_e$ is a given convergence accuracy.
For the small-timescale optimization algorithm, the most computationally intensive operation lies in the matrix inversion for updating $\mathbf{w}$ in \eqref{w_update2}, which is $\mathcal{O} (M^3)$.
However, this operation only needs to run once within each short time duration $T_c$ because the inverse of the matrix remains constant in different iterations.
Finding the optimal dual variable $\pi^{\star}$ in \eqref{u_update_final} only needs to solve a cubic equation, and the associated complexity is negligible.
So, the overall complexity within a long time duration $T_s$ is $ \mathcal{O} \left(I \big( M^3 + MN + N\log(\frac{1}{\epsilon_e}) \big) + \frac{T_s}{T_c} M^3  \right)$, where $I$ denotes the iteration number of the large-timescale CSSCA algorithm.

\section{Simulation Results}
In this section, we validate the effectiveness of the two-timescale optimization scheme for the IRS-assisted D2D underlay communication system. 
Consider a three-dimensional (3D) coordinate system, the BS and the IRS are respectively located at $[12,0,3]^T$ and $ [0,50,3]^T $ with a uniform linear array (ULA), where $M = 8$ and $N = 60$.
The CU, the DT and the DR are laid in a semi-circle around the IRS, whose coordinates are $[12,50,0]^T$, $[2,60,0]^T$, and $[2,40,0]^T$, respectively.
The path loss of each channel is modeled by $L = L_0 \left( d/d_0 \right)^{-\alpha}$, where $L_0 = -30$ dB denotes the reference path loss at the reference distance $d_0 = 1$ m. $d$ denotes the link distance, and $\alpha$ denotes the path loss exponent.
For the BS-IRS link, the BS-CU link, and the BS-DR link, the path loss exponents are set as $\alpha_{\rm bi} = 2.2$ and $\alpha_{\rm bu} = \alpha_{\rm br} = 3.6$, respectively; $\alpha_{\rm iu}, \alpha_{\rm ti}, \alpha_{\rm ir}, \alpha_{\rm tu},$ and $ \alpha_{\rm tr}$ are set to be $2.8$.
For small-scale fading, the Rician factor of the BS-IRS link is set as $\kappa_{\rm bi} = 20$ dB. 
Due to severe direct path blockage, the Rician factors of the BS-CU link and the BS-DR link are set as $\kappa_{\rm bu} = \kappa_{\rm br} = 0$ (i.e., Rayleigh fading). The Rician factors of other links, $\kappa_{\rm iu}, \kappa_{\rm ti}, \kappa_{\rm ir}, \kappa_{\rm tu},$ and $\kappa_{\rm tr} $, are set to be $\kappa$.
The remaining parameters are set as: $p_0 = 10$ dBm, $\sigma^2_{\rm u} = \sigma^2_{\rm r} = -80$ dBm, $\Gamma_{\rm u} = 12$ dB, $\epsilon = 0.05$, $\beta = 10^{3}$, $\tau_0 = \tau_1 = 0.005$, $\rho^t = t^{-0.8}$, $\gamma^t = t^{-1}$, $\frac{T_s}{T_c}=200$.

For better comparison, we consider an upper bound and three benchmark schemes as follows:

\begin{itemize}
	\item \textit{Instantaneous CSI:}
	$\boldsymbol{\varphi}$, $\mathbf{w}$ and $p$ are jointly optimized in each time duration of $T_c$ based on the instantaneous CSI.
	\item \textit{Statistical CSI:}
	$\boldsymbol{\varphi}$, $\mathbf{w}$ and $p$ are jointly optimized in the time duration of $T_s$ based on the statistical channel mean.
	\item \textit{Random phase shifts:}
	$\mathbf{w}$ and $p$ are jointly optimized in each time duration of $T_c$  based on the instantaneous CSI; while $\boldsymbol{\varphi}$ is randomly set.
	\item \textit{Without IRS:}
	$\mathbf{w}$ and $p$ are jointly optimized in each time duration of $T_c$ based on the instantaneous CSI without the existence of IRS.
\end{itemize}

\begin{figure} 
	[!ht]
	\centering
	\vspace{-1.0em}
	\includegraphics[width=0.83\columnwidth]{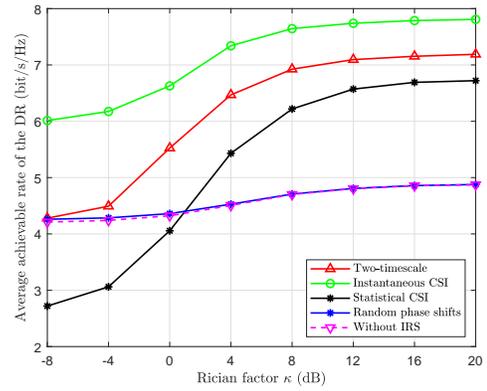}
	\caption{Average achievable rate of the DR versus $\kappa$, with $p_1 = 1$ dBm.}
	\label{RicianFactor}
	\vspace{-0.2em}
\end{figure}
In Fig. \ref{RicianFactor}, the average achievable rate of the DR is evaluated over the Rician factor $\kappa$ when the maximum transmit power of the DT is $p_1 = 1$ dBm.
One can see that, the proposed two-timescale algorithm outperforms the \textit{random phase shifts} and the \textit{without IRS} schemes significantly especially when $\kappa$ is large (about 50\% increase when $\kappa = 20$ dB). 
This is because as $\kappa$ increases, the related channels become more deterministic, and a larger proportion of the slow-varying LoS components can be obtained to boost the performance of the proposed two-timescale optimization scheme.
It is also observed that, the performance gap between the \textit{instantaneous CSI} scheme and the proposed scheme diminishes with the increase of $\kappa$, which seems to approach a constant eventually.
The constant gap is expected since both the BS-CU link and the BS-DR link hold Rayleigh fading, no LoS component can be extracted in the above two links to further improve the performance. 
Compared with the \textit{statistical CSI} scheme, the performance improvement of the proposed scheme is due to the instantaneous CSI based small-timescale optimization gain.

\begin{figure} 
	[!ht]
	\centering
	\vspace{-1.0em}
	\includegraphics[width=0.83\columnwidth]{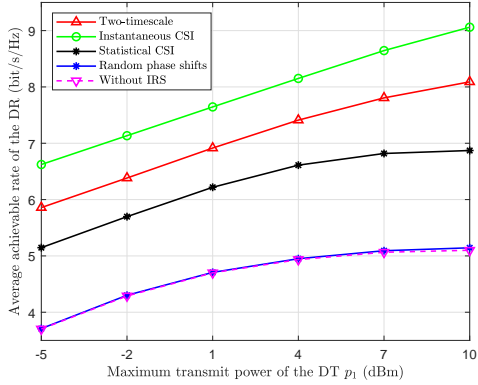}
	\caption{Average achievable rate of the DR versus $p_1$, with $\kappa = 8$ dB.}
	\label{Power}
	\vspace{-0.2em}
\end{figure}
Fig. \ref{Power} shows the average achievable rate of the DR versus the maximum transmit power of the DT $p_1$, with $\kappa = 8$ dB.
It is evident that the average achievable rates of all schemes increase with $p_1$.
However, the curves tend to become saturated when $p_1$ is large, which is more obvious in the \textit{random phase shifts} and the \textit{without IRS} schemes.
This is because the DT power $p$ is also restricted by the SINR outage constraint of the CU.
It is also observed that compared to the \textit{instantaneous CSI} scheme, the proposed scheme only incurs about $10\%$ average achievable rate degradation.
This demonstrates an attractive trade-off between system performance and CSI overhead.

\section{Conclusion}
In this paper, we presented a two-timescale optimization scheme to study the IRS-assisted D2D underlay system.
An efficient algorithm was established to optimize the BS beamformer, the IRS phase shifts, and the DT power based on different kinds of CSI.
Numerical results demonstrated that the proposed scheme is attractive compared to three baselines, and also embraces high implementation prospect.

\appendix
The first-order derivative of the surrogate function $g_0\left(\boldsymbol{\varphi},\mathbf{w},p; \mathbf{H} \right)$ w.r.t. $\boldsymbol{\varphi}$ is given by
\begin{equation}
\nabla_{\boldsymbol{\varphi}}g_0\left(\boldsymbol{\varphi},\mathbf{w},p; \mathbf{H} \right) = \frac{2}{\ln 2} \Re \left\{ -j \boldsymbol{\theta}^* \circ \left( \frac{a_0}{\Gamma_0}  -
\frac{a_1}{\Gamma_1} \right) \right\} ,
\end{equation}
where $\circ$ denotes the Hadamard product, and
\begin{equation}
\Gamma_{0} = \left| \left( \boldsymbol{\theta}^H \mathrm{diag} \left( \mathbf{h}_{\rm ir} \right) \mathbf{G}_{\rm bi} + \mathbf{h}_{\rm br} \right) \mathbf{w}\right|^{2} + {\sigma}_{\rm r}^{2},
\end{equation}
\begin{equation}
\begin{aligned}
\Gamma_{1} = \Gamma_{0} + p \left| \boldsymbol{\theta}^H \mathrm{diag} \left( \mathbf{h}_{\rm ir} \right) \mathbf{h}_{\rm ti} + {h}_{\rm tr} \right| ^{2},
\end{aligned}
\end{equation}
%and
\begin{equation}
\begin{aligned}
a_{0} = \mathrm{diag} \left( \mathbf{h}_{\rm ir} \right) \mathbf{G}_{\rm bi} \mathbf{w} \mathbf{w}^{H} \left(\mathbf{G}_{\rm bi}^H \mathrm{diag} \left( \mathbf{h}_{\rm ir}^H \right)  \boldsymbol{\theta} + \mathbf{h}_{\rm br}^H \right),
\end{aligned}
\end{equation}
\begin{equation}
\begin{aligned}
a_{1} = a_{0} + p \mathrm{diag} \left( \mathbf{h}_{\rm ir} \right) \mathbf{h}_{\rm ti}  \left( \mathbf{h}_{\rm ti}^H \mathrm{diag} \left( \mathbf{h}_{\rm ir}^H \right)  \boldsymbol{\theta} +
{h}_{\rm tr}^* \right).
\end{aligned}
\end{equation}

Similarly, the first-order derivative of the surrogate function $g_1\left(\boldsymbol{\varphi},\mathbf{w},p; \mathbf{H} \right)$ w.r.t. $\boldsymbol{\varphi}$ is given by
\begin{equation}
\nabla_{\boldsymbol{\varphi}}g_1\left(\boldsymbol{\varphi},\mathbf{w},p; \mathbf{H} \right) = 2 \hat{u}_{\beta}^{'} \left( Q \right) \Re \left\{ -j \boldsymbol{\theta}^* \circ \nabla_{\boldsymbol{\varphi}}Q \right\} ,
\end{equation}
where
\begin{equation}
\hat{u}_{\beta}^{'} \left( Q \right) = \frac{\beta e^{-\beta Q}}{(1+ e^{-\beta Q})^2} ,
\end{equation}
%and
\begin{align}
\nabla_{\boldsymbol{\varphi}}Q =& \Gamma_{\rm u} p \mathrm{diag} \left( \mathbf{h}_{\rm iu} \right) \mathbf{h}_{\rm ti} \left( \mathbf{h}_{\rm ti}^H \mathrm{diag} \left( \mathbf{h}_{\rm iu}^H \right)  \boldsymbol{\theta} + {h}_{\rm tu}^*\right) \nonumber - \\
& \mathrm{diag} \left( \mathbf{h}_{\rm iu} \right) \mathbf{G}_{\rm bi} \mathbf{w} \mathbf{w}^{H}  (\mathbf{G}_{\rm bi}^H \mathrm{diag} \left( \mathbf{h}_{\rm iu}^H \right) \boldsymbol{\theta} + \mathbf{h}_{\rm bu}^H).
\end{align}

\bibliographystyle{IEEEtran}%By using IEEEtrans, the number can be displayed.
\bibliography{IEEEabrv,mybib}

\end{document}